\documentclass[%
 aip,
 jmp,%
 amsmath,
 amssymb,
 reprint,%
 nofootinbib
]{revtex4-1}

\usepackage{graphicx}
\usepackage{dcolumn}
\usepackage{bm}
\usepackage{bbm}
\usepackage{natbib}
\usepackage[section]{algorithm}
\usepackage{algorithmic}
\usepackage{verbatim}
\usepackage{amsthm}
\usepackage{multirow}
\usepackage{epstopdf}

\theoremstyle{definition}
\newtheorem{algo}{Algorithm}
\newcommand{\q}{\bm{q}}
\newcommand{\p}{\bm{p}}

\newcommand{\z}{\bm{z}}
\newcommand{\D}{\bm{D}}
\newcommand{\F}{\bm{F}}
\newcommand{\dt}{{\Delta t}}
\newcommand{\ds}{{\Delta s}}
\newcommand{\R}{\bm{R}}
\newcommand{\M}{\bm{M}}
\newcommand{\bPhi}{\bm{\Phi}}

\begin{document}


\title[]{Variable length trajectory compressible hybrid Monte Carlo}

\author{Akihiko Nishimura}
 \email{an88@duke.edu.}
 \affiliation{\mbox{Department of Mathematics, Duke University, Durham, North Carolina, 27708, USA}}
\author{David Dunson}%
\affiliation{\mbox{Department of Statistical Science, Duke University, Durham, North Carolina, 27708, USA}}%

\date{\today}

\begin{abstract}
\begin{changemargin}{0cm}{8ex}
	Hybrid Monte Carlo (HMC) generates samples from a prescribed probability distribution in a configuration space  by simulating Hamiltonian dynamics, followed by the Metropolis (-Hastings) acceptance/rejection step. Compressible HMC (CHMC) generalizes HMC to a situation in which the dynamics is reversible but not necessarily Hamiltonian. This article presents a framework to further extend the algorithm. Within the existing framework, each trajectory of the dynamics must be integrated for the same amount of (random) time to generate a valid Metropolis proposal.  Our generalized acceptance/rejection mechanism allows a more deliberate choice of the integration time for each trajectory. The proposed algorithm in particular enables an effective application of variable step size integrators to HMC-type sampling algorithms based on reversible dynamics. The potential of our framework is further demonstrated by another extension of HMC which reduces the wasted computations due to unstable numerical approximations and corresponding rejected proposals.
	
\end{changemargin}
	
\end{abstract}

                             

\maketitle

\section{\label{sec:intro}Introduction}
	A study of molecular systems often relies on generating random variables from a prescribed (unnormalized) probability distribution $\rho(\q) \propto \exp(-U(\q))$ on the configuration space. Markov chain Monte Carlo (MCMC) generates samples from a target distribution by constructing a Markov chain whose stationary distribution coincides with the target distribution. Such a Markov chain can be realized by building a transition rule that satisfies the detailed balance condition. MCMC based on the Metropolis (-Hastings) algorithm \cite{metropolis53} is a general sampling approach widely used in computational physical science as well as in Bayesian statistics and machine learning. Many such algorithms are inefficient, producing highly correlated samples, and require a large number of iterations to adequately characterize the target distribution \cite{roberts01, mattingly12, pillai12}. 
	
	
	Hybrid Monte Carlo \cite{duane87} (HMC) constructs the proposal distribution for the Metropolis algorithm by simulating molecular dynamics (MD), a procedure that can greatly reduce the correlation among successive MCMC samples. More precisely, HMC augments the state space by introducing a \textit{momentum} variable $\p$; the original variable $\q$ is often referred to as \textit{position} variable in the HMC framework. In this augmented state space $(\q,\p)$, the proposal distribution for Metropolis is constructed by solving an ordinary differential equation (ODE) corresponding to Newton's equations of motion with respect to the potential energy $U(\q)$. There are applications in which (partial) analytical solutions to an ODE can be exploited \cite{shahbaba13, pakman13, pakman14}, but in general ODEs are discretized and integrated numerically. 
	
	Within the original HMC framework, an integrator for simulating MD must be reversible and volume-preserving to produce a valid Metropolis proposal \cite{neal10}. In fact, the volume-preserving property can be relaxed by including a Jacobian factor in the calculation of the Metropolis acceptance probability. \cite{leimkuhler09, lan15} Under this generalization of HMC, any reversible (discrete) dynamics / bijective map can be applied to generate a proposal state. This algorithm is formalized as \textit{compressible HMC} (CHMC) in Ref.~\onlinecite{fang14}. A generalization of HMC known as \textit{Riemannian manifold HMC} (RMHMC) \cite{girolami11} also falls within the framework of CHMC.
	
	This article presents an algorithm to relax another condition required by (compressible) HMC. Given a reversible map $\bm{F}$ and state $(\q,\p)$, CHMC proposes the next state by applying the map $n$ times, where the number of steps $n$ can be drawn randomly at each iteration. Though often not stated explicitly, the detailed balance requires the number of steps to be determined independently of the trajectory $\{(\q,\p), \bm{F}(\q,\p), \bm{F}^2(\q,\p), \ldots \}$. As we will show in Section~\ref{sec:vlt_chmc}, this constraint can prevent realizing the full potential of MCMC algorithms based on reversible dynamics. 
	
	Our algorithm generalizes the acceptance-rejection mechanism behind CHMC to allow the number of steps to depend on each trajectory of the dynamics while preserving the detailed balance. The number of numerical integration steps taken in simulating a trajectory of HMC is commonly referred to as the ``path length'' of a trajectory in the statistics literature. We therefore call our algorithm \textit{variable length trajectory CHMC} (VLT-CHMC). It should be mentioned that the No-U-Turn-Sampler (NUTS) is another variant of HMC that allows the path lengths to vary from one trajectory to another. \cite{hoffman14} However, the motivation behind NUTS is to spare a user the trouble of manually tuning the number of steps, and NUTS in general performs no better than HMC with well-chosen path lengths. \cite{hoffman14, wang13} On the other hand VLT-CHMC can improve the performance of CHMC in a more fundamental and significant way. In particular, VLT-CHMC enables an effective application of reversible variable step size integrators to HMC-type sampling algorithms based on reversible dynamics. 
	
	The rest of the paper is organized as follows. Section~\ref{sec:chmc} reviews the main ideas behind CHMC and provides an example in which the compressible dynamics arises from the use of non-traditional integrators in HMC settings. Such integrators have proven to be more efficient than the commonly used volume-preserving integrators in various applications.  The example also serves to introduce the notations and concepts needed in the next section, where VLT-CHMC is motivated as a method to effectively apply variable step size integrators in HMC settings. The presentation is self-contained, but some familiarity with HMC is assumed. VLT-CHMC is developed in Section~\ref{sec:vlt_chmc}. Section~\ref{sec:vlt_chmc_special} explains how the existing framework limits the utility of variable step size integrators to sampling algorithms. The key observation in addressing this issue leads to a special case of VLT-CHMC. More general construction of VLT-CHMC is provided in Section~\ref{sec:vlt_chmc_general}. Section~\ref{sec:rahmc} presents another use case of VLT-CHMC, where HMC is modified to reduce the wasted computation due to unstable numerical approximations and corresponding rejected proposals. The simulation results are shown in Section~\ref{sec:simulation} to demonstrate the potential gains from the framework of VLT-CHMC.
	
	
\section{Review of compressible HMC}
\label{sec:chmc}

\subsection{Basic Theory}
	To keep the description of CHMC and the subsequent development of VLT-CHMC more intuitive, the version of CHMC described here is slightly less general than the one in Ref.~\onlinecite{fang14}. It is straightforward to extend the variable length trajectory algorithm of Section~\ref{sec:vlt_chmc} to the general settings.
	
	A bijective map $\F$ is said to be \textit{reversible} if
	\begin{equation}
	\label{eq:reversibility}
	\F^{-1} = \R \circ \F \circ \R
	\end{equation}
or equivalently $\left( \R \circ \F \right)^{-1} = \R \circ \F$ for an involution $\R$ (i.e.\ $\R \circ \R = \text{id}$). Note that the reversiblity of $\F$ implies that of $\F^n$ for any $n$. Let $\D \F^n$ denote the Jacobian matrix of $\F^n$ and $|\D \F^n|$ its determinant. Given a state $\z$ and integer $n$, CHMC proposes the state $\z^* = \R \circ \F^n (\z)$ and accepts or rejects the proposal with probability
	\begin{equation}
	\label{eq:acceptance_prob}
	\min \left\{ 1, 
		\frac{ \rho(\z^*) | \D \F^n(\z) | }{ \rho(\z)} \right\}
	\end{equation}
	To see that this transition rule satisfies the detailed balance with respect to $\rho(\cdot)$, consider a small neighborhood $B$ around $\z$ and $B^* = \R \circ \F^n(B)$ around $\z^*$, so that $\R \circ \F^n(B^*) = B$. The proposal move sends the probability mass
	\[ \int_B \rho(\z') {\rm d}\z' \approx \rho(\z) \text{vol}(B) \]
	from $B$ to $B^*$. On the other hand, the mass sent from $B^*$ to $B$ by the proposal move can be seen to be
	\begin{align*}
	\int_{B^*} \rho(\z') {\rm d}\z' 
		&= \int_{B} \rho(\z') |\D (\R \circ \F^n)(\z')|{\rm d}\z' \\
		&\approx \rho(\z^*) |\D \F^n(\z)| \, \text{vol}(B)
	\end{align*}
	by the change of variable formula and the fact $|\R| = 1$. The acceptance and rejection step of CHMC amounts to rejecting the fraction of move by the ratio of the probability fluxes and thus imposes the detailed balance.

	The above transition rule preserves the target density $\rho(\cdot)$ for any $n$, so in practice the number of steps can be drawn randomly at each iteration of CHMC.  The steps of CHMC are summarized in Algorithm~\ref{alg:chmc} below, where the distribution $p(\cdot)$ for the number of steps is a tuning parameter a user must specify.  The use of a deterministic map as a proposal distribution does not yield an ergodic Markov chain, and therefore such a transition rule must be alternated with another transition rule that preserves the target density $\rho(\cdot)$, as done in Step~1 of the algorithm. We do not concern ourselves here with how to choose such a random move since the choice depends critically on the particular form of $\rho(\cdot)$.
\begin{algo}[Compressible HMC]
	\label{alg:chmc}
	With a prespecified probability mass function $p(\cdot)$ on $\mathbb{Z}^+$, CHMC generates a Markov chain $\{ \z^{(m)} \}_m$ with the following transition rule $\z^{(m)} \to \z^{(m+1)}$:
	\begin{enumerate}
	\item Make a random change $\z^{(m)} \to \z$ that preserves the target density $\rho(\cdot)$.
	\item Sample $n \sim p(\cdot)$ and propose the state $\z^* = \R \circ \F^n(\z)$.
	\item Let $\z^{(m+1)} = \z^*$ with probability
   	  \[ \min \left\{1, 
   		\frac{ \rho(\z^*) | \D \F^n(\z) | 
   		}{ 
   		\rho(\z)} \right\} \]
   		Otherwise, let $\z^{(m+1)} = \z$.
	\end{enumerate}
\end{algo}

\subsection{Example: (Riemann manifold) HMC with non-volume-preserving integrators}
\label{sec:hmc_wo_volume_preservation}
	HMC and its extension Riemann manifold HMC (RMHMC) construct a reversible and volume-preserving bijective map by numerically approximating Hamiltonian dynamics. To this end, they require a geometric integrator that preserves the reversibility and volume-preservation property of Hamiltonian dynamics. Under the CHMC framework, however, Hamiltonian dynamics can be approximated using a wider range of integration techniques. 
	
	
	In order to sample from a probability density of interest $\rho_0(\q) \propto \exp(-U(\q))$ in $\mathbb{R}^d$, RMHMC introduces an auxiliary variable $\p \in \mathbb{R}^d$ whose distribution is defined conditionally as $\p | \q \sim \mathcal{N}(\bm{0}, \M(\q))$ for a family of positive definite matrices known as \textit{mass tensors}  $\{\M(\q)\}_{\q}$. \cite{bennett75, neal10, girolami11}  The joint density $\rho(\q,\p)$ in the phase space then is given as $\rho(\q,\p) \propto \exp(-H(\q,\p))$ where the \textit{Hamiltonian} $H(\q,\p)$ is given by
	\begin{equation}
	\label{eq:rmhmc_hamiltonian}
	H(\q,\p)
		= U(\q) + \frac{1}{2} \p^T \M(\q)^{-1} \p
			+ \frac{1}{2} \log | \M(\q) |
	\end{equation}
	The proposal is generated by approximating the solution to \textit{Hamilton's equations}:
	\begin{equation}
	\label{eq:hamilton's}
	\frac{\text{d} \q}{\text{d} t}
		= \nabla_{\p} H(\q, \p), \
	\frac{\text{d} \p}{\text{d} t}
		= - \nabla_{\q} H(\q, \p)
	\end{equation}
	For the Hamiltonian \eqref{eq:rmhmc_hamiltonian}, the solution operator of \eqref{eq:hamilton's} is reversible with respect to a momentum flip operator $\R(\q,\p) = (\q,-\p)$. Solving \eqref{eq:hamilton's} using a reversible integrator with a constant step size $\dt$ yields a reversible map $\F_{\dt}$ so that 
	\begin{equation}
	\label{eq:approximate_map}
	\F_{\dt}^n(\q_0,\p_0) \approx (\q(n \dt), \p(n \dt))
	\end{equation}
	where $\{(\q(t), \p(t))\}_t$ denotes the exact solution with the initial condition $(\q_0,\p_0)$. In other words, $\F_\dt$ approximates the solution operator $\bPhi_\dt$ of \eqref{eq:hamilton's} defined through the relation
	\begin{equation}
	\label{eq:solution_operator}
	\frac{\text{d} \bPhi_t}{\text{d} t}
		= \big( (\nabla_{\p} H) \circ \bPhi_t, - (\nabla_{\q} H) \circ \bPhi_t \big)
	\end{equation}
	for all $t$. If the reversible map $\F_{\dt}$ is further required to be volume preserving, then we have $|\D \F_{\dt}^n| = 1$ and the Jacobian factor drops from \eqref{eq:acceptance_prob}, recovering HMC and RMHMC algorithms of Ref.~\onlinecite{duane87, girolami11}. In some applications however, non-volume-preserving approximations of \eqref{eq:hamilton's} have been shown to offer substantial gains in computational efficiency. \cite{lan15, fang14} 
	
	For example, Lan et.\ al.~\cite{lan15} considers the ODE corresponding to \eqref{eq:hamilton's} in terms of reparametrization $(\q, \mathbf{v}) = (\q, \M(\q)^{-1} \p)$.  The reparametrized ODE admits semi-explicit and explicit reversible approximations, requiring fewer or no fixed point iterations compared to the St\"{o}rmer-Verlet integrator typically employed in RMHMC. The proposal move using a simulated trajectory is alternated with sampling $\mathbf{v}$ from its conditional density $\mathbf{v} | \q \sim \mathcal{N}(\bm{0}, \M(\q)^{-1})$, a random move corresponding to Step 1 in Algorithm~\ref{alg:chmc}.
	The CHMC algorithm based on the semi-explicit and explicit integrator are found to significantly outperform RMHMC based on the St\"{o}rmer-Verlet integrator over a range of examples.

\section{Variable length trajectory CHMC}
\label{sec:vlt_chmc}
	Variable length trajectory CHMC (VLT-CHMC) is most naturally motivated as a method to effectively apply variable step size integrators in RMHMC settings. For this reason, we first develop this special case of VLT-CHMC in Section~\ref{sec:vlt_chmc_special}. A more general theory is developed in Section~\ref{sec:vlt_chmc_general}. Section~\ref{sec:rahmc} illustrates the use and potential benefits of the general VLT-CHMC algorithm through another example.
	

\subsection{Special case of VLT-CHMC}
\label{sec:vlt_chmc_special}

\subsubsection{Motivation: RMHMC with variable step size integrators and limitations of CHMC}
	In Section~\ref{sec:hmc_wo_volume_preservation}, we discussed how CHMC allows us to approximate Hamiltonian dynamics with non-volume-preserving integrators and still generate a valid Metropolis proposal. We in particular considered the use of a reversible integrator with a constant step size. A wider range of reversible integration techniques for Hamiltonian systems are available in the literature, however, including a number of variable step size integrators. \cite{calvo98, blanes12, leimkuhler04, hairer06}	In theory, a variable step size integrator similarly produces a valid CHMC proposal as long as the integrator is reversible. However, the use of such an integrator under the existing CHMC framework generally leads to an algorithm with suboptimal sampling efficiency, for the reasons we describe now.
	
	Each step of a variable step size integrator approximates the evolution $(\q(t_n), \p(t_n)) \to (\q(t_n + \dt_n), \p(t_n + \dt_n))$ where the step size $\dt_n$ depends on the current state $(\q(t_n), \p(t_n))$ through a \textit{step size controller} $g(\q,\p)$. The simplest choice of step size would be $\dt_n =  g(\q(t_n),\p(t_n)) \ds$, but the reversibility requires a slightly more sophisticated relationship and the condition $g(\q,\p) = g(\q,-\p)$ (see Section~\ref{sec:vlt_chmc_algorithm_in_special_case}).  
	Most importantly for our discussion, a variable step size scheme is equivalent to approximating the following \textit{time-rescaled} Hamiltonian dynamics in a new time scale $\text{d} s = g(\q,\p)^{-1} \text{d} t$ with a constant step size $\ds$:
	\begin{equation}
	\label{eq:time_rescaled_hamilton}
	\frac{\text{d} \q}{\text{d} s}
		= g(\q, \p) \nabla_{\p} H(\q, \p), \
	\frac{\text{d} \p}{\text{d} s}
		= - g(\q, \p) \nabla_{\q} H(\q, \p)
	\end{equation}
	In other words, a reversible variable step size approximation of \eqref{eq:hamilton's} yields a reversible map $\F_{\ds}$ such that 
	\begin{equation}
	\label{eq:approximate_time_rescaled_map}
	\F_{\ds}^n(\q_0,\p_0) \approx (\q(n \ds), \p(n \ds)) 
	\end{equation}
	where $\{\q(s),\p(s)\}_s$ is the solution to the time-rescaled dynamics \eqref{eq:time_rescaled_hamilton} with the initial condition $(\q_0,\p_0)$. 
	
	This implicit time-rescaling behind variable step size integration causes trouble for CHMC. The utility of Hamiltonian dynamics \eqref{eq:hamilton's} as a proposal generation mechanism stems from the fact that $\rho(\q,\p) \propto \exp(-H(\q,\p))$ is the \textit{invariant distribution} of the dynamics i.e.\ if $(\q_0,\p_0)$ has the distribution $\rho(\q,\p) \propto \exp(-H(\q,\p))$, then $\bPhi_t(\q_0,\p_0)$ also has the same distribution $\rho(\cdot)$ for all $t$.
	As a consequence, the proposal generated by an approximate solution $(\q^*,\p^*) = \F_{\dt}^n(\q_0, \p_0)$ as in \eqref{eq:approximate_map} can be accepted with probability 1 in the limit $\dt \to 0$ and $n \dt \to t'$. On the other hand, the time-rescaled dynamics \eqref{eq:time_rescaled_hamilton} in general does not preserve the target density $\rho(\q,\p)$, and the proposal generated by the approximate solution $(\q^*, \p^*) = \F_{\ds}^n(\q_0,\p_0)$ may not be accepted with high probability even in the limit $\ds \to 0$ and $n \ds \to s'$. In fact, the acceptance probability of the CHMC proposal in the limit is given by:
	\begin{equation}
	\label{eq:variable_stepsize_chmc_acceptance_prob}
	\min \left\{ 1, \frac{ g(\q(s'),\p(s')) }{ g(\q_0,\p_0) } \right\}
	\end{equation}
	where $\{\q(s),\p(s)\}_s$ denotes the solution to \eqref{eq:time_rescaled_hamilton} with the initial condition $(\q_0,\p_0)$. The derivation is given in Appendix~\ref{app:acceptance_prob_derivation}.
	

\subsubsection{Algorithm: variable length trajectory scheme for time-rescaled dynamics}
\label{sec:vlt_chmc_algorithm_in_special_case}
	In order to address the issue caused by the implicit time-rescaling associated with variable step size integrators, VLT-CHMC approximates the dynamics in the original time scale as follows. Fix the initial condition $(\q_0, \p_0)$ and denote $(\q_i,\p_i) = \F_\ds^i (\q_0,\p_0)$ where $\F_\ds$ approximates the dynamics in the time scale $s$ as in \eqref{eq:approximate_time_rescaled_map}. The evolution $(\q_0, \p_0) \to (\q(t),\p(t))$ in the original time scale can be approximated by taking the trajectory dependent number of steps $N(\q_0, \p_0) = N(t, \q_0, \p_0)$ defined as
	\begin{equation}
	\label{eq:traj_len_function}
	\begin{aligned}
	N(& \q_0, \p_0) = \\
		&\min \left\{ n :
		\sum_{i=1}^n \frac{\ds}{2} \left( g(\q_{i - 1}, \p_{i - 1}) + g(\q_{i}, \p_{i}) \right) > t \right\}
	\end{aligned}
	\end{equation}
	Now we consider the map $\F_\ds^N$ defined as
	\begin{equation}
	\F_\ds^N (\q, \p) = \F_\ds^{N(\q,\p)} (\q,\p)
	\end{equation}
	which approximates the solution operator $\bPhi_t$ as defined in \eqref{eq:solution_operator}. The map however cannot be used directly to generate a proposal because in general it is neither reversible or even bijective. The map would be reversible if $N(\q_0^*, \p_0^*) = N(\q_0, \p_0)$ where $(\q_0^*, \p_0^*) = \R \circ \F_\ds^N (\q_0, \p_0)$, but \eqref{eq:traj_len_function} only implies $N(\q_0^*, \p_0^*) \leq N(\q_0, \p_0)$. For example when $g(\q_0^*, \p_0^*) \gg g(\q_0, \p_0)$, the simulated time along the reverse trajectory $\left\{ (\q_i^*, \p_i^*) = \F_\ds^i(\q_0^*, \p_0^*) \right\}_{i=0}^n$ 
	\[ \sum_{i=1}^n \frac{\ds}{2} \big( g(\q^*_{i - 1}, \p^*_{i - 1}) + g(\q^*_{i}, \p^*_{i}) \big) \]
	will likely reach the threshold $t$ before $n = N(\q_0, \p_0)$ steps.
	
	The key observation behind VLT-CHMC is that we can nonetheless construct collections of states $S$ and $S^*$ containing $(\q_0, \p_0)$ and $(\q_0^*, \p_0^*)$ such that
	\begin{equation}
	\label{eq:generalized_reversibility}
	\begin{aligned}
	\R \circ \F_\ds^N(S) \subset S^* 
		&\ \text{ and } \
		\R \circ \F_\ds^N(S^*) \subset S \\
	\R \circ \F_\ds^N(S^c) \subset (S^{*})^c 
		&\ \text{ and } \
		\R \circ \F_\ds^N \left((S^{*})^c \right) \subset S^c
	\end{aligned}
	\end{equation}
	The existence of such sets $S$ and $S^*$ is a property of the map $\F_\ds^N$ and generalizes the notion of reversibility \eqref{eq:reversibility}.
	The set $S$ is essentially the pre-image of $\{(\q_0^*, \p_0^*)\}$ under $\R \circ \F_\ds^N$ and can be constructed by defining $S = \left\{(\q_{- \ell}, \p_{- \ell}), \thinspace \ldots, (\q_r, \p_r) \right\}$ by choosing $\ell, r \geq 0$ such that
	\begin{equation} 
	\label{eq:nstep_forward_backward}
	\begin{aligned}
	\ell
		&= \max \left\{ j \geq 0: \F_\ds^N(\q_{- j}, \p_{- j}) = \F_\ds^N(\q_0, \p_0) \right\} \\
	r
		&= \max \left\{ j \geq 0: \F_\ds^N(\q_{j}, \p_{j}) = \F_\ds^N(\q_0, \p_0) \right\}
	\end{aligned}
	\end{equation}
	Algorithmically, $\ell$ and $r$ can be found by solving the dynamics backward and forward from $(\q_0, \p_0)$ using the equivalent definitions below:	
	\begin{equation}
	\label{eq:nstep_special_case}
	\begin{aligned}
	\ell
		&=\max \left\{ j \geq 0: \sum_{i=-j}^{N(t,\q_0,\p_0) - 1}  \dt_i < t \right\} \\
	r
		&= \max \left\{ j \geq 0:  \sum_{i=j}^{N(t,\q_0,\p_0)} \dt_i > t \right\} \\
		&\hspace{3ex} \text{ where } \dt_i = \frac{\ds}{2} \left( g(\q_{i - 1}, \p_{i - 1}) + g(\q_{i}, \p_{i}) \right)
	\end{aligned}
	\end{equation}
	The set $S^*$ is the pre-image of $\{(\q_r, \p_r)\}$ under $\R \circ \F_\ds^N$ and can analogously be constructed. Denoting $(\q^*_i, \p^*_i) = \F_\ds^i(\q_0^*, \p_0^*)$, let $S^* = \left\{(\q^*_{- \ell^*}, \p^*_{- \ell^*}), \thinspace \ldots, (\q^*_{r^*}, \p^*_{r^*}) \right\}$ where $\ell^*, r^* \geq 0$ is defined as 
	\begin{equation} 
	\label{eq:nstep_forward_backward_star}
	\begin{aligned}
	\ell^*
		&= \max \left\{ j \geq 0: \F_\ds^N(\q^*_{- j}, \p^*_{- j}) = \F_\ds^N(\q^*_0, \p^*_0) \right\} \\
	r^*
		&= \max \left\{ j \geq 0: \F_\ds^N(\q^*_{j}, \p^*_{j}) = \F_\ds^N(\q^*_0, \p^*_0) \right\}
	\end{aligned}
	\end{equation}
	It is shown in Appendix~\ref{app:vlt_chmc_justification} that the above definition actually implies $r^* = 0$. The proof of \eqref{eq:generalized_reversibility} and of other facts regarding $S$ and $S^*$ are also given in Appendix~\ref{app:vlt_chmc_justification}.

	Having constructed the sets $S$ and $S^*$ with the property \eqref{eq:generalized_reversibility}, VLT-CHMC imposes the detailed balance by rejecting a fraction of moves between $S$ and $S^*$ as described in Algorithm~\ref{alg:vlt_chmc} below.
	\begin{algo}[VLT-CHMC]
	\label{alg:vlt_chmc}
	Given a reversible map $\F_\ds$ as in \eqref{eq:approximate_time_rescaled_map} and a trajectory length function $N$ as in \eqref{eq:traj_len_function}, VLT-CHMC generates a Markov chain $\{(\q^{(m)}, \p^{(m)})\}_m$ with the following transition rule $(\q^{(m)}, \p^{(m)}) \to (\q^{(m+1)}, \p^{(m+1)})$:
	\begin{enumerate}
	\item Sample $\p_0$ from the conditional density $\p | \q^{(m)}$ and set $\q_0 = \q^{(m)}$. 
	\item Find the indices $\ell, r, \ell^*, r^*$ as in \eqref{eq:nstep_forward_backward} and \eqref{eq:nstep_forward_backward_star} by simulating the dynamics forward and backward from $(\q_0, \p_0)$ and $\R \circ \F_\ds^N(\q_0,\p_0)$. Then set
	\begin{align*}
	S &= \left\{ \F_\ds^{- \ell}(\q_0, \p_0), \thinspace \ldots, \F_\ds^r(\q_0, \p_0) \right\} \\
	S^* &= \Big\{ \R \circ \F_\ds^{N_0 - r^*}(\q_0, \p_0), \\
		&\hspace{15ex} \thinspace \ldots, \R \circ \F_\ds^{N_0 + \ell^*}(\q_0, \p_0) \Big\}
	\end{align*}
	where $N_0 = N(\q_0, \p_0)$.
	\item Propose the transition from $S$ to $S^*$ with the acceptance probability which is the smaller of 1 and 
		\begin{equation} 
		\label{eq:valet_accept_prob}		
			\frac{
			\sum\limits_{j=-r*}^{\ell^*} \rho \left( \R \circ \F_\ds^{N_0+j}(\q_0, \p_0) \right) \left|\D \F_\ds^{N_0+j}(\q_{0},\p_{0}) \right|
			}{
			\sum\limits_{i=-\ell}^r \rho \left( \F_\ds^i(\q_0, \p_0) \right) \left| \D \F_\ds^{i}(\q_{0},\p_{0}) \right|
			} 
		\end{equation}
	\item If the transition in Step~3 is accepted, choose a state $\R \circ \F_\ds^{N_0+j}(\q_0, \p_0)$ from $S^*$ with the probability proportional to
		\begin{equation}
			\rho \left( \R \circ \F_\ds^{N_0+j}(\q_0, \p_0) \right) \left|\D \F_\ds^{N_0+j}(\q_{0},\p_{0}) \right|
		\end{equation}
		and set $(\q^{(m+1)}, \p^{(m+1)}) = \R \circ \F_\ds^{N_0+j}(\q_0, \p_0)$. Otherwise, choose a state $\F_\ds^i(\q_0, \p_0)$ from $S$ with the probability proportional to
		\begin{equation}
		\rho \left( \F_\ds^i(\q_0, \p_0) \right) \left| \D \F_\ds^{i}(\q_{0},\p_{0}) \right|
		\end{equation}
		and set $(\q^{(m+1)}, \p^{(m+1)}) = \F_\ds^i(\q_0, \p_0)$.
	\end{enumerate}
	\end{algo}

\subsubsection{Theory: VLT-CHMC and detailed-balance condition}
\label{sec:detailed_balance_for_vlt_chmc}
	Too see how VLT-CHMC achieves the detailed balance, consider a small neighborhood $B_0$ around $(\q_0, \p_0)$. The total probability in the neighborhood $B = \cup_{i=- \ell}^r \F_\ds^i(B_0)$ of $S$ is
	\begin{equation}
	\label{eq:mass_around_S}
	\begin{aligned}
	\int_{B} 
		&\rho(\q,\p) \, {\rm d} \q \, {\rm d} \p \\
		&\approx \sum_{i=- \ell}^r \rho \left( \F_\ds^i(\q_0, \p_0) \right) \left| \D \F_\ds^{i}(\q_{0},\p_{0}) \right| \big|B_0 \big|
	\end{aligned}
	\end{equation}
	assuming that $B_0$ is small enough that $\F_\ds^i(B_0)$'s are disjoint. Similarly, the total probability in the neighborhood $B^* = \cup_{j=- r^*}^{\ell^*} \R \circ \F_\ds^{N_0+j}(B_0)$ of $S^*$ is 
	\begin{equation}
	\label{eq:mass_around_S_star}
	\begin{aligned}
	\int_{B^*} 
		&\rho(\q,\p) \, {\rm d} \q \, {\rm d} \p \\
		&\hspace{-2ex} \approx \sum_{j=- r^*}^{\ell^*} \rho \left( \R \circ \F_\ds^{N_0+j}(\q_0, \p_0) \right) \left|\D \F_\ds^{N_0+j}(\q_{0},\p_{0}) \right| \big|B_0 \big|
	\end{aligned}
	\end{equation}
	Comparing the acceptance probability \eqref{eq:valet_accept_prob} with the probability fluxes \eqref{eq:mass_around_S} and \eqref{eq:mass_around_S_star}, one can see that the acceptance-rejection procedure of Step~3 controls the probability fluxes appropriately to achieve the detailed balance between the neighborhoods $B$ and $B^*$. Step~4 then imposes the detailed balance within $B$ and $B^*$ by sampling a state according to the relative amount of probability in the individual components $\{ \F_\ds^i(B_0) \}_{i=- \ell}^r$ of $B$ and $\{ \R \circ \F_\ds^{N_0+j}(B_0) \}_{j=- r^*}^{\ell^*}$ of $B^*$.
	
\subsubsection{Theoretical efficiency: improvement over CHMC}
	Throughout Section~\ref{sec:vlt_chmc_special} we considered the compressible dynamics \eqref{eq:time_rescaled_hamilton} arising from a variable step size integration of Hamiltonian dynamics. In this specific setting with the trajectory length function $N$ as defined in \eqref{eq:traj_len_function}, VLT-CHMC is guaranteed to have a high average acceptance probability. In fact, in the limit $\ds \to 0$ with $t$ fixed, the acceptance probability \eqref{eq:valet_accept_prob} of a VLT-CHMC proposal from $(\q_0,\p_0)$ converges to a value bounded below by
	\begin{equation}
	\label{eq:vlt_accept_prob_lower_bound}
	\frac{ g(\bPhi_t(\q_0,\p_0)) }{ g(\q_0,\p_0) } 
		\left \lfloor \frac{ g(\q_0,\p_0) }{ g(\bPhi_t(\q_0,\p_0)) } \right \rfloor
	\end{equation}
	when $g(\bPhi_t(\q_0,\p_0)) < g(\q_0,\p_0)$. In case $g(\bPhi_t(\q_0,\p_0)) > g(\q_0,\p_0)$, a similar lower bound holds for the proposal from $\R \circ \bPhi_t(\q_0,\p_0)$. Note that the quantity \eqref{eq:vlt_accept_prob_lower_bound} is always larger than $1/2$ and it tends to 1 as the ratio $g(\bPhi_t(\q_0,\p_0)) / g(\q_0,\p_0)$ increases, in contrast with the acceptance probability \eqref{eq:variable_stepsize_chmc_acceptance_prob} of CHMC. More precise results on the acceptance probability of a VLT-CHMC proposal are derived in Appendix~\ref{app:acceptance_prob_derivation}.
	
	Of course, the acceptance rate of a proposal distribution is not the only factor determining the efficiency of an MCMC algorithm. Nonetheless, the theoretical result above highlights an advantage VLT-CHMC has over the usual CHMC. The bottom line is that VLT-CHMC proposals approximate the original dynamic \eqref{eq:hamilton's} while CHMC proposals approximate the time-rescaled dynamics \eqref{eq:time_rescaled_hamilton}. Therefore, VLT-CHMC will generally outperform CHMC whenever the exact solution of the original dynamics constitutes an efficient Markov chain propagator as is typically the case in RMHMC applications. \cite{girolami11, nishimura16} This is substantiated by our simulation study in Section~\ref{sec:simulation}.

\subsection{General VLT-CHMC}
\label{sec:vlt_chmc_general}
	The key step in Algorithm~\ref{alg:vlt_chmc} is the construction of the sets $S$ and $S^*$ with the property \eqref{eq:generalized_reversibility}. More generally, the detailed balance can be imposed by the same type of acceptance-rejection mechanism whenever the phase space can be partitioned into a collection of pairs $S$ and $S^*$ such that the set $S \cup S^*$ and $(S \cup S^*)^c$ is closed under a (deterministic) transition rule. Conceivably, a wide range of algorithms can be devised under this general condition. In this section we present one systematic way to generalize the framework of Section~\ref{sec:vlt_chmc_special}.
	
	Consider a generic reversible map $\F$ on a state space $\z$ and associated involution $\R$. Fix $\z_0$ and denote $\z_i = \F^i(\z_0)$. Choose a \textit{trajectory termination criteria}, or more precisely boolean valued functions $b_n(\z_0, \ldots, \z_n) \in \{0, 1\}$, with the following property
	\begin{equation}
	\label{eq:criteria_symmetry}
	b_n(\z_0, \ldots, \z_n) = b_n(\R(\z_n), \ldots, \R(\z_0))
	\end{equation}
	as well as the property
	\begin{equation}
	\label{eq:criteria_monotonicity}
	b_n(\z_0, \ldots, \z_n) = 1 \ \text{ only if } \ 
		b_{n-i}(\z_i, \ldots, \z_n) = 1
	\end{equation}
	for any $i > 0$. These properties are satisfied, for example, by a termination criteria $\sum_{i=1}^n a(\z_{i}) + a(\z_{i-1}) > c$ for a scalar function $a(\z) \geq 0$. Define a corresponding trajectory length function $N(\z_0)$ as
	\begin{equation}
	\label{eq:traj_length_function_general}
	\begin{aligned}
	N(\z_0) 
		&= \min \{ N'(\z_0), N_{\rm max} \} \\
		&\hspace{-2ex} \text{for } \ N'(\z_0) = \min \big\{ n :  b_n(\z_0, \ldots, \z_n) = 1
		\big\} 
	\end{aligned}
	\end{equation}
	With the reversible map $\F_\ds$ and trajectory length function $N$ of \eqref{eq:traj_len_function} replaced by the generic ones as above, Algorithm~\ref{alg:vlt_chmc} remains a valid MCMC scheme. This is because the justification of the algorithm (in Appendix~\ref{app:vlt_chmc_justification}) only require a trajectory length function $N$ to satisfy the \textit{short return} condition
	\begin{equation}
	\label{eq:short_return}
	N(\z^*) \leq N(\z) \ \text{ where } \z^* = \R \circ \F^{N(\z)} (\z)
	\end{equation}
	and \textit{order preserving} condition
	\begin{equation}
	\label{eq:order_perserving}
	N(\z) - n \leq N(\F^n(\z)) \ \text{ for any } n
	\end{equation}
	The intuition behind the terminologies are explained in Appendix~\ref{app:vlt_chmc_justification} along with the proof of the general VLT-CHMC algorithm.
	

\subsection{Example: Rejection Avoiding HMC}
\label{sec:rahmc}
	Here we illustrate a use of the general VLT-CHMC framework through an algorithm of very different flavor from the special case presented in Section~\ref{sec:vlt_chmc_special}. 
	
	A step size required for stable numerical integration of Hamilton's equation \eqref{eq:hamilton's} can vary significantly at different regions of a phase space in some application areas of HMC. \cite{neal10} In such situations, the Hamiltonian may be approximately preserved along a simulated trajectory for a while until it suddenly starts to deviate wildly, leading to a proposal with little chance of acceptance. VLT-CHMC provides a way to ``detect'' when the trajectory becomes unstable and select an alternate state along the trajectory to transition to.
	
	Let $\F_\dt$ be a volume-preserving and reversible map as in \eqref{eq:approximate_map}, approximating Hamiltonian dynamics. Consider a trajectory $\left\{(\q_i, \p_i) = \F_\dt^i (\q_0, \p_0) \right\}_{i = 0, 1, 2, \ldots}$. When the trajectory becomes unstable, it can be detected by a trajectory termination criteria such as
	\begin{equation}
	\label{eq:max_H_fluctuation_criteria}
	b_n = \mathbbm{1} \left\{ \max_{0 \leq i \leq n} H(\q_i, \p_i) - \min_{0 \leq i \leq n} H(\q_i, \p_i)
		\geq \epsilon \right\}
	\end{equation}
	where $\mathbbm{1}$ is an indicator function. We will actually use an alternative criteria below since this leads to a simpler algorithm implementation:
	\begin{equation}
	\label{eq:error_per_step_criteria}
	\begin{aligned}
	b_n = \mathbbm{1} \Big\{ 
		& \left| H(\q_i, \p_i) - H(\q_{i-1}, \p_{i-1}) \right| \geq \epsilon \\
		& \hspace{15ex} \text{ for some } \, i = 1, \ldots, n \Big\}
	\end{aligned} 
	\end{equation}
	It is easy to check that the criteria \eqref{eq:max_H_fluctuation_criteria} and \eqref{eq:error_per_step_criteria} satisfy the properties \eqref{eq:criteria_symmetry} and \eqref{eq:criteria_monotonicity} and define a valid trajectory length function $N$ of the form \eqref{eq:traj_length_function_general} for Algorithm~\ref{alg:vlt_chmc}. We refer to the version of VLT-CHMC based on the criteria \eqref{eq:error_per_step_criteria} as \textit{rejection avoiding HMC}.
	
	A proposal of rejection avoiding HMC recovers the usual HMC proposal with the trajectory length $N_{\rm max}$ when the fluctuation of a Hamiltonian at each step is within the error tolerance $\epsilon$. However, upon detecting the fluctuation of magnitude larger than $\epsilon$ at the step $(\q_{i-1}, \p_{i-1}) \to (\q_i, \p_i)$, the algorithm proceeds to simulate the trajectory backward from $(\q_0, \p_0)$ and $(\q_0^*, \p_0^*) = (\q_i, - \p_i)$ to determine the sets $S$ and $S^*$ according to the rule in Step~2 of Algorithm~\ref{alg:vlt_chmc}. 
	

\section{Numerical Results}
\label{sec:simulation}

\subsection{Geometrically tempered HMC with variable step size integrator}
	HMC is known to have a serious difficulty sampling from a multi-modal target density as the potential energy barriers among the modes prevents transition from one mode to another. To address this issue, Nishimura and Dunson \cite{nishimura16} propose a version of RMHMC with a mass tensor having the property
	\begin{equation}
	\label{eq:gthmc_mass_tensor}
	| \M(\q) |^{1/2}
		\propto \rho(\q)^{1 - T^{- 1}}
	\end{equation}
	with a \textit{temperature} parameter $T \geq 1$.  It can be shown that, with such a choice of a mass tensor, RMHMC algorithm is equivalent to the usual HMC algorithm (with a constant mass tensor) applied to a tempered distribution $\tilde{\rho}(\tilde{\q}) \propto \rho(\q)^{1/T}$ on a manifold parametrized by $\tilde{\q}$. For this reason, RMHMC with the property \eqref{eq:gthmc_mass_tensor} is referred to as \textit{geometrically tempered HMC} (GTHMC) in Ref.~\onlinecite{nishimura16}.  
	
	The typical velocity of the dynamics \eqref{eq:hamilton's} at the position $\q$ is given by the operator norm $\lVert \M(\q) \rVert^{- 1/2}$. This quantity, and in turn the velocity of the dynamics, necessarily becomes unboundedly large in the regions where $\rho(\q)$ is small, due to the constraint \eqref{eq:gthmc_mass_tensor}. For this reason, the only practical way to approximate the dynamics underlying GTHMC algorithms is through a variable step size integrator with a step size proportional to $\lVert \M(\q) \rVert^{1/2}$. 
	
	We take an example with a simple bimodal target density from Ref.~\onlinecite{nishimura16}. The density $\rho(\q)$ is defined as a mixture of two-dimensional Gaussians with unit-variance centered at $(4, 0)$ and $(-4, 0)$. The mass tensor is chosen as	
	\begin{equation}
	\label{eq:dthmc_mass_tensor}
	\begin{aligned}
	\M(\q) 
		&\propto \rho(\q)^{2 \gamma \left( 1 - T^{- 1} \right) } \bm{e}_1 \bm{e}_1^T \\
		&\qquad + \rho(\q)^{2 \left( 1-\gamma \right) (d-1)^{- 1} \left( 1 - T^{- 1} \right) } \left( \bm{I} - \bm{e}_1 \bm{e}_1^T \right)
	\end{aligned}
	\end{equation}
	for $d^{-1} \leq \gamma \leq 1$ where $d = 2$ is the dimension of $\q$ and $\bm{e}_1 = (1,0)$ is a standard basis vector. 
	The mass tensors suggested in Ref.~\onlinecite{lan14, fang14} have apparent resemblance to \eqref{eq:dthmc_mass_tensor}, but the crucial difference is that they do not satisfy \eqref{eq:gthmc_mass_tensor} and consequently offer rather limited improvement over the standard HMC.
	
	We compare the performance of CHMC and VLT-CHMC with the explicit variable step size integrator developed in Ref.~\onlinecite{nishimura16}. VLT-CHMC is run with the trajectory length function \eqref{eq:traj_len_function}.  The main challenge in this example to explore the phase space along the first coordinate of $\q$ due to the multi-modality along this direction. Therefore the efficiency of the sampling algorithms is summarized by the effective sample sizes (ESS) along the first coordinate of $\q$. The ESS's as well as the acceptance probabilities at different parameter settings of CHMC and VLT-CHMC are summarized in Table~\ref{table:bimodal_results_chmc} and \ref{table:bimodal_results_vlt_chmc}. As predicted by our discussion in Section~\ref{sec:vlt_chmc_special}, VLT-CHMC has substantially higher acceptance probabilities and, across various parameter settings, is five times more efficient than CHMC with the optimal parameter choice.  The time step size $\ds = .75$ for the variable step size integrator was used for all the simulations and was chosen to control the error in the Hamiltonian within a reasonable level along the trajectories. ESS's were computed using the initial monotone sequence estimator of Geyer. \cite{geyer92}
	

\begin{table}                                                               
\centering                                                                        
\begin{tabular}{cccccccc}                                                 
\hline                                                                            
Number of steps & 5 & 10 & 15 & 20 & 25 & 30 & 35 \\                              
\hline                                                                            
Acceptance rate & 0.48 & 0.38 & 0.37 & 0.36 & 0.34 & 0.33 & 0.33 \\         
\hline                                                                            
ESS & 75.7 & 180 & 145 & 83.1 & 103 & 123 & 101 \\       
\hline                                                                            
\end{tabular}                                                                     
\caption{ESS of CHMC along the first coordinate per $10^5$ force evaluations at the various numbers of numerical integration steps. The number of steps coincides with that of force evaluations.}
\label{table:bimodal_results_chmc}                                                
\end{table}  

\begin{table}                                                                  
\centering                                                                           
\begin{tabular}{cccccccc}                                                    
\hline                                                                               
$t$ & 0.50 & 0.75 & 1.00 & 1.25 & 1.50 & 1.75 & 2.00 \\
\hline
Number of steps & 13 & 17 & 21 & 24 & 27 & 30 & 33 \\            
\hline                                                                               
Acceptance rate & 0.81 & 0.78 & 0.76 & 0.75 & 0.73 & 0.72 & 0.71 \\              
\hline                                                                               
ESS & 899 & 966 & 924 & 992 & 925 & 921 & 805 \\            
\hline                                                                               
\end{tabular}                                                                        
\caption{ESS of VLT-CHMC along the first coordinate per $10^5$ force evaluations. The integration time $t$ determines the trajectory lengths through the termination criteria in \eqref{eq:traj_len_function}. }
\label{table:bimodal_results_vlt_chmc}                                               
\end{table}

\subsection{Rejection avoiding HMC}
\begin{figure}
	\centering
	\includegraphics[width=1\linewidth]{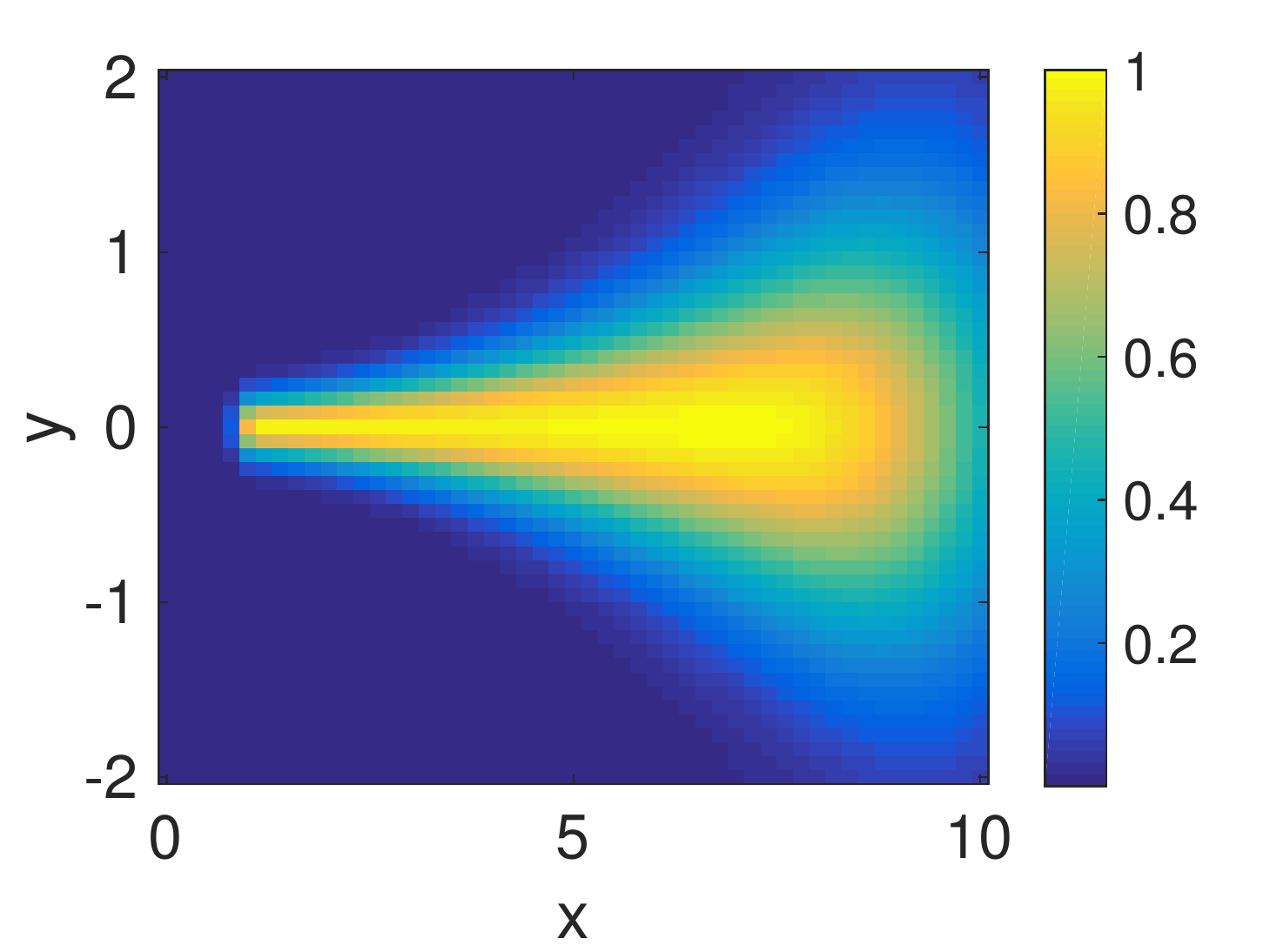} 
	\caption{A plot of (unnormalized) probability density function $\rho(x,y) \propto \exp( - U(x,y))$ used to illustrate the benefit of rejection avoiding HMC.}
	\label{fig:club_shape_density}
	\end{figure}

\begin{figure}
	\centering
	\includegraphics[width=.9\linewidth]{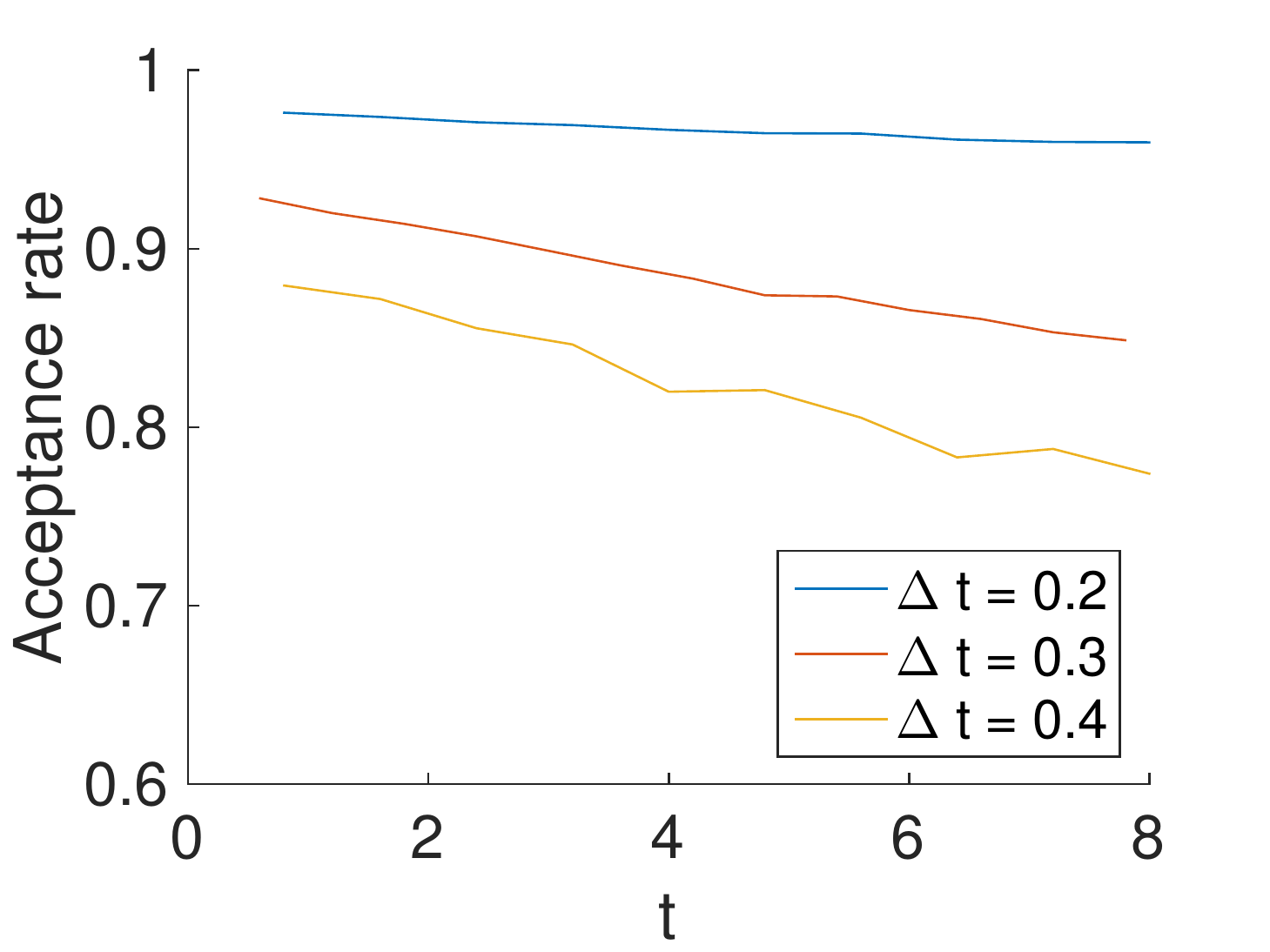} 
	\caption{Acceptance rate of HMC proposals at various settings of step size and integration time when sampling from the density shown in Figure~\ref{fig:club_shape_density}.}
	\label{fig:club_shape_acceptance_rate}
	\end{figure}

\begin{figure}
	\centering
	\includegraphics[width=.9\linewidth]{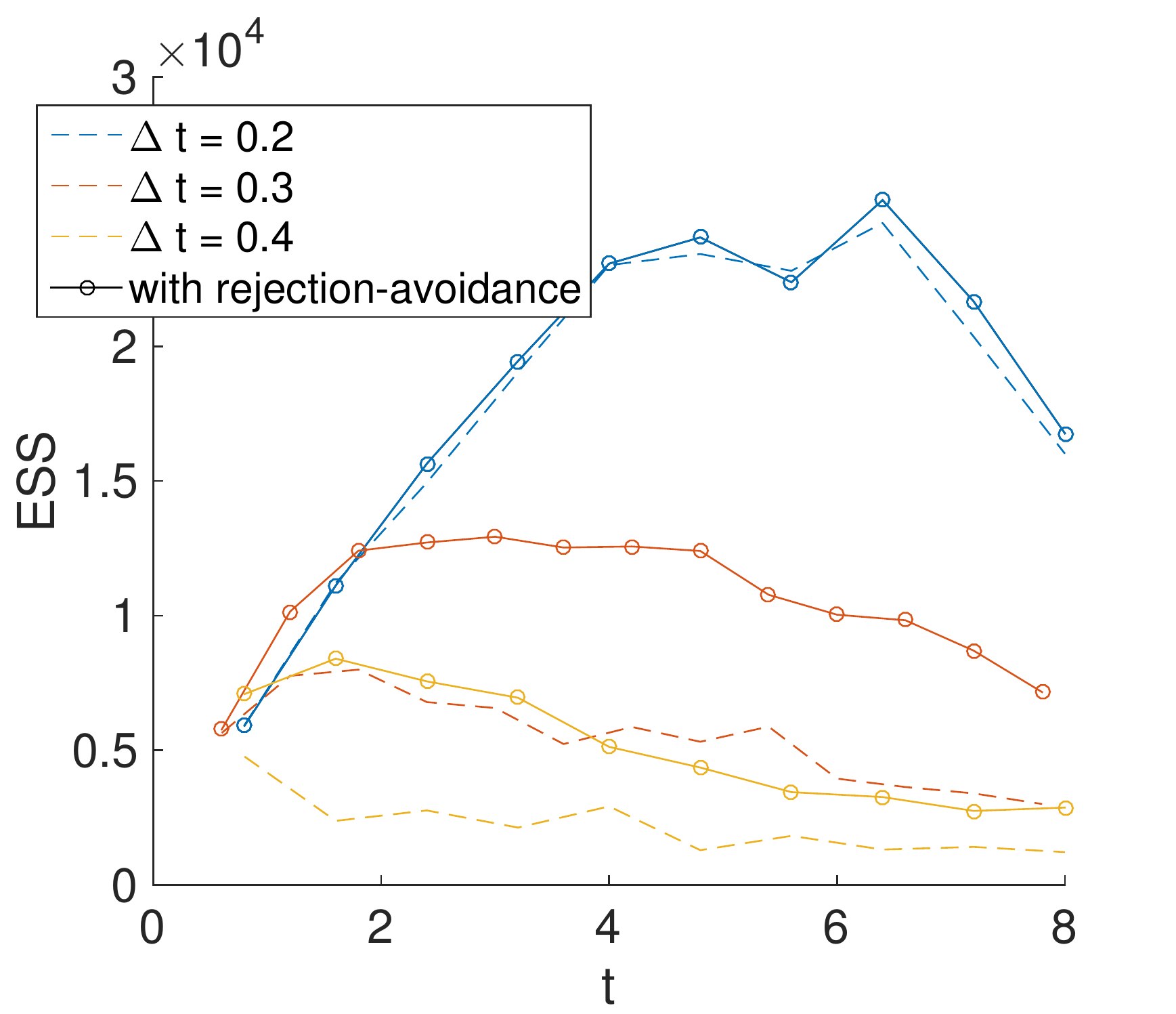} 
	\caption{ESS per $10^6$ force evaluations at various settings of step size and integration time. The ESS's are for the mean estimation along the $x$-axis. }
	\label{fig:club_shape_ess}
	\end{figure}

	To illustrate the benefit of the rejection avoiding algorithm described in Section~\ref{sec:rahmc}, we consider the problem of sampling from a probability density function $\rho(x,y) \propto \exp( - U(x,y))$ as plotted in Figure~\ref{fig:club_shape_density}. The density $\rho(x,y)$ is constructed as a (continuous) Gaussian mixture
	\begin{equation}
	\rho(x,y) \propto
		\int_1^{10} \frac{1}{\sigma_\mu} \exp \left(- \frac{(x-\mu)^2}{2 \, \sigma_\mu^2} - y^2 \right) {\rm d} \mu
	\end{equation}
	where $\sigma_\mu = 0.1 + (\mu / 10)^2$.  The density has a property that, along the $x$-axis, the partial derivative $\partial_y U(x,y)$ varies substantially and so does the stable step size for the leap-frog integrator typically employed in HMC.  For example, the leap-frog integrator with the step size $\dt \geq 0.4$ approximates the Newton's equations of motion quite accurately in the region $x > 4$, while the step size of $\dt \approx 0.2$ is required for a numerically stable approximation in the region $x < 2$. In in practice, such a knowledge is obviously not available to us and the appropriate step size must be determined empirically from preliminary runs of HMC.  A common strategy is to pick a target acceptance rate for the HMC proposals, typically in the range $0.65 \sim 0.8$, and tune the step size accordingly. \cite{beskos13, neal10, stan15}  This approach would suggest a step size well above the stability in this example, however.  Figure~\ref{fig:club_shape_acceptance_rate} shows that the acceptance rate of HMC to be quite high even for the step size $\dt = 0.4$.  The acceptance rate can be high despite some unstable trajectories because the region where the approximation become unstable contains relatively small, though not negligible, probability.  On the other hand, the performance of HMC is severely undermined by the choice of a too large step size as can be seen in Figure~\ref{fig:club_shape_ess}.  The ESS's for $10^6$ force evaluations, estimated from ten independent simulations, are shown so that the computational cost is fixed across the experiments. The error tolerance in Hamiltonian, as in \eqref{eq:error_per_step_criteria}, for rejection avoiding HMC is set to $\epsilon = 3$.  When $\dt = 0.2$, less than 1\% of trajectories experience the error in Hamiltonian above the tolerance, so there is no practical difference between HMC with and without rejection avoidance.  However, without rejection avoidance, the ESS is reduced by the factor as large as five when increasing the step size from $\dt = 0.2$ to $\dt = 0.3$. The performance degradation is less severe for rejection avoiding HMC as the algorithm concentrates the computational efforts on the stable portions of approximated trajectories.  
	
	In summary, choosing an optimal step size for HMC is difficult in practice as the choice must be made without the detailed knowledge of a target density. A step size can appear to approximate the dynamics accurately but be above the stability limit in some regions. Rejection avoiding HMC can alleviate the effect of a suboptimal step size choice and provides far more ESS's than the standard HMC in such situations.
		
\section{Acknowledgments}
	We would like to thank Jiangfeng Lu for his feedback on a preliminary draft of the manuscript.

\appendix

\section{Derivation of limiting acceptance probability} 
\label{app:acceptance_prob_derivation}
	In this section we analyse the acceptance probability of CHMC and VLT-CHMC algorithms in the special case of RMHMC with variable step size integrators as described in Section~\ref{sec:vlt_chmc_special}. We derive explicit formulas as well as useful bounds on the acceptance probabilities in the limit $\ds \to 0$.

\newcommand{\Abs}[1]{\left| #1 \right|}
\newcommand{\bphi}{\bPhi}
\subsection{Acceptance probability of CHMC}
\label{app:accept_prob_of_chmc}
	
	When approximating a time-rescaled Hamiltonian dynamics \eqref{eq:time_rescaled_hamilton} with a reversible map $\F_\ds$ as in \eqref{eq:approximate_time_rescaled_map}, the acceptance probability of the CHMC proposal from $(\q,\p)$ is calculated by the formula
	\[ 1 \wedge \frac{\rho(\R \circ \F_\ds^n(\q,\p)) \left| {\D \F_\ds^n(\q,\p)} \right| }{\rho(\q,\p)}  \]
In the limit $\ds \to 0$ and $n \ds \to {s'}$, the above quantity converges to
	\[ 1 \wedge \frac{\rho(\R \circ \bPhi_{s'}(\q,\p)) \left| \D \bPhi_{s'}(\q,\p) \right|}{\rho(\q,\p)} \]
where $\bPhi_{s}$ is the solution operator of the dynamics \eqref{eq:time_rescaled_hamilton} i.e.\ 
	\begin{equation}
	\frac{\text{d} \bPhi_s}{\text{d} s}
		= (g \circ \bPhi_s) \ \mathbf{f} \circ \bPhi_s
	\end{equation}
	where $\mathbf{f} = (\nabla_{\p} H, - \nabla_{\q} H)$. We have $\rho \circ \bPhi_{s'} = \rho$ since Hamiltonian dynamics conserves the energy and so does the time-rescaled dynamics. We also have $\rho \circ \R = \rho$, so that $\rho(\R \circ \bPhi_{s'}(\q,\p)) = \rho(\q,\p)$. To establish the limiting acceptance probability \eqref{eq:variable_stepsize_chmc_acceptance_prob}, therefore, it remains to show that $\left| \D \bPhi_{s'}(\q,\p) \right| = g(\bPhi_{s'}(\q,\p)) / g(\q,\p)$. The Jacobian $\D \bPhi_s$ satisfies a matrix-valued differential equation $\frac{\partial}{\partial s} \D \bPhi_s = \D \mathbf{f} \circ \bPhi_s \, \D \bPhi_s$ and therefore Liouville's formula tells us that
	\begin{align*}
	\left| \D \bPhi_{s'} \right|
		&=  \exp \left( \int_0^{s'} {\rm tr} \left( \D \mathbf{f} \circ \bPhi_s \right) {\rm d} s \right)
	\end{align*}
	A straightforward calculation shows that ${\rm tr} \left( \D \mathbf{f} \circ \bPhi_s \right) = \frac{\partial}{\partial s} \log g \circ \bPhi_s$, from which the identity $\left| \D \bPhi_{s'} \right| = g \circ \bPhi_{s'} / g$ follows.

\subsection{Acceptance probability of VLT-CHMC}
\label{app:accept_prob_of_vlt_chmc}
	In the derivation below, we will follow the notations of Section~\ref{sec:vlt_chmc_algorithm_in_special_case}. Namely, we set $(\q_0^*, \p_0^*) = \R \circ \F^N(\q_0,\p_0)$, $(\q_i, \p_i) = \F_\ds^i(\q_0, \p_0)$, and $(\q_i^*, \p_i^*) = \F_\ds^i(\q_0^*, \p_0^*)$. The trajectory length function $N = N(t)$ is defined as in \eqref{eq:traj_len_function} and the sets $S$ and $S^*$ as in Algorithm~\ref{alg:vlt_chmc}. Note that $(\q_0,\p_0)$ is fixed, but other quantities depend on $\ds$, including but not limited to $(\q_i, \p_i)$'s, $N(\q_0, \p_0)$, and $S$. We do not denote the dependence explicitly but it is implied.	
	
We will show that the acceptance probability of the transition from $S$ to $S^*$ converges to 
	\begin{equation}
	\label{eq:accept_prob_limit_vltchmc}
	1 \wedge \frac{g(\bPhi_t(\q_0, \p_0)) |S^*|}{g(\q_0,\p_0) |S|}
	\end{equation}
	 as $\ds \to 0$ while $t$ fixed. Moreover, if $g(\bPhi_t(\q_0, \p_0)) < g(\q_0, \p_0)$, then in the limit $\ds \to 0$ we have $|S| = 1$ and
	\begin{equation}
	\label{eq:set_size_vlt_chmc}
	\frac{g(\q_0,\p_0)}{g(\bPhi_t(\q_0, \p_0))} - 1
		\leq \left| S^* \right|
		\leq \frac{g(\q_0,\p_0)}{g(\bPhi_t(\q_0, \p_0))} + 1
	\end{equation}
	The claimed lower bound \eqref{eq:vlt_accept_prob_lower_bound} on the acceptance probability follows immediately from \eqref{eq:accept_prob_limit_vltchmc} and \eqref{eq:set_size_vlt_chmc}.
	
	It is not difficult to show that ${\rm diam}(S) \to 0$ and ${\rm diam}(S^*) \to 0$ as $\ds \to 0$. This means that the elements of $S$ (and of $S^*$) collapse to a single state as $\ds \to 0$. More precisely, for all $- \ell \leq i \leq r$ and $- r^* \leq j \leq \ell^*$,
	\begin{equation}
	\label{eq:set_size_collapsing}	
	\begin{aligned}
	\F_\ds^i(\q_0,\p_0) 
		&\to (\q_0,\p_0)  \\
	\R \circ \F_\ds^{N_0+j}(\q_0,\p_0)
		&\to \R \circ \bPhi_t(\q_0,\p_0)
	\end{aligned}
	\end{equation}
	where $N_0 = N(\q_0, \p_0)$ and $r, \ell, r^*, \ell^*$ are defined as in \eqref{eq:nstep_forward_backward} and \eqref{eq:nstep_forward_backward_star}. It follows that	
	\begin{equation} \label{subeq:denom_num_limit}
	\begin{aligned}
	&\rho \left(\F_\ds^i(\q_0,\p_0) \right) \Abs{\D\F_\ds^{i}(\q_{0},\p_{0})} 
		\to \rho(\q_0,\p_0) \\
	&\rho \left(\R \circ \F_\ds^{N_0+j}(\q_0,\p_0) \right) \left| \D \F_\ds^{N_0+j}(\q_0,\p_0) \right| \\
	&\hspace{5em} \to \rho(\R \circ \bPhi_t(\q_0,\p_0)) \left| \D \bPhi_t(\q_0,\p_0)) \right|
	\end{aligned}
	\end{equation}	
	By the same argument as in Section~\ref{app:accept_prob_of_chmc}, we can show that 
	\begin{equation}
	\begin{aligned}
	\rho(\R \circ \bPhi_t(\q_0,\p_0)) 
		&\left| \D \bPhi_t(\q_0,\p_0)) \right| \\
		&= \rho(\q_0, \p_0) \frac{g(\bPhi_t(\q_0,\p_0))}{g(\q_0,\p_0)}
	\end{aligned}
	\end{equation}
	establishing the claimed formula \eqref{eq:accept_prob_limit_vltchmc}.

	We now turn to the proof of the inequality \eqref{eq:set_size_vlt_chmc}. The intuition behind the inequality and the proof below is that the size of the set $\Abs{S^*}$ is roughly equal to the number of intervals of length $\ds \cdot g(\q_0^*,\p_0^*)$ that can be fit inside the interval $\left( t, \, t + \ds \cdot g(\q_0,\p_0) \right)$. Denote $N_0^* = N(\q_0^*, \p_0^*)$. By the definition of $N_0^*$, $r^*$, and $\ell^*$, we must have
	\begin{equation}
	\begin{aligned}
	 &\sum_{i=-\ell^*+1}^{N_0^*-1}\frac{\ds}{2} \left(g(\q_{i-1}^*,\p_{i-1}^*) + g(\q_i^*,\p_i^*)\right) \\
	 	&\hspace{5ex} < t 
		< \sum_{i=r^*+1}^{N_0^*} \frac{\ds}{2} \left(g(\q_{i-1}^*,\p_{i-1}^*) + g(\q_i^*,\p_i^*)\right)
	\end{aligned}
	\end{equation}
	which implies that 
	\begin{equation} 
	\label{eq:ineq1_for_valet_accept_prob}
	\begin{aligned}
	&\sum_{i=-\ell^*+1}^{r^*} \frac{1}{2} \left(g(\q_{i-1}^*,\p_{i-1}^*) + g(\q_i^*,\p_i^*)\right) \\
		&\hspace{8ex} < \frac{1}{2} \left(g(\q_{N_0^*-1}^*,\p_{N_0^*-1}^*) + g(\q_{N_0^*}^*,\p_{N_0^*}^*) \right)
	\end{aligned}
	\end{equation}
	Also by the definition $N_0^*$, $r^*$, and $\ell^*$, we must have
	\begin{equation}
	\begin{aligned}
	&\sum_{i=r^*+2}^{N_0^*}\frac{\ds}{2} \left(g(\q_{i-1}^*,\p_{i-1}^*) + g(\q_i^*,\p_i^*)\right) \\
		&\hspace{6ex} < t 
		< \sum_{i=-\ell^*}^{N_0^*-1} \frac{\ds}{2} \left(g(\q_{i-1}^*,\p_{i-1}^*) + g(\q_i^*,\p_i^*)\right)
	\end{aligned}
	\end{equation}
	which implies that
	\begin{equation} 
	\label{eq:ineq2_for_valet_accept_prob}
	\begin{aligned}
	&\frac{1}{2} \left(g(\q_{N_0^*-1}^*,\p_{N_0^*-1}^*) + g(\q_{N_0^*}^*,\p_{N_0^*}^*) \right) \\
		&\hspace{8ex} < \sum_{i=-\ell^*}^{r*+1} \frac12 \left(g(\q_{i-1}^*,\p_{i-1}^*) + g(\q_i^*,\p_i^*)\right)
	\end{aligned}
	\end{equation}
	Since ${\rm diam}(S) \to 0$ and ${\rm diam}(S^*) \to 0$ as $\ds \to 0$, the inequalities \eqref{eq:ineq1_for_valet_accept_prob} and \eqref{eq:ineq2_for_valet_accept_prob} converge to
	\begin{equation} \label{subeq:setsize_bounds}
	g(\q_0^*,\p_0^*) (|S^*|-1) 
		\leq g(\q_0,\p_0) 
		\leq g(\q_0^*,\p_0^*) (|S^*| + 1)
	\end{equation}
	The desired inequality \eqref{eq:set_size_vlt_chmc} is obtained by rearranging the terms in the above inequality. 
	
	Finally, we turn to the proof of the fact that $\Abs{S} \to 1$ as $\ds \to 0$ when $g(\q_0^*,\p_0^*) < g(\q_0,\p_0)$. To this end, we only need to note that all the arguments in the proof of \eqref{subeq:setsize_bounds} remain valid if we switch the role of $(\q_i^*,\p_i^*)$, $r^*$, $\ell^*$ and $N_0^*$ with $(\q_i,\p_i)$, $r$, $\ell$ and $N_0$. This means that the inequality \eqref{subeq:setsize_bounds} still holds if we switch the role of $S^*$ with $S$ and of $(\q_0^*,\p_0^*)$ with $(\q_0,\p_0)$, yielding the inequality
	\[ g(\q_0,\p_0) (\Abs{S}-1) 
		\leq g(\q_0^*,\p_0^*) 
		\leq g(\q_0,\p_0) (\Abs{S} + 1) \]
	In particular, we have $\Abs{S}	\leq \frac{g(\q_0^*,\p_0^*)}{g(\q_0,\p_0)} + 1$ and hence $\Abs{S} = 1$.

\section{Justification of VLT-CHMC algorithm}
\label{app:vlt_chmc_justification}
	As claimed in Section~\eqref{sec:vlt_chmc_general}, Algorithm~\ref{alg:vlt_chmc} remains a valid algorithm when we replace the reversible map $\F_\ds$ with any reversible map and the trajectory length function $N$ with any function of the form \eqref{eq:traj_length_function_general}. In Section~\ref{sec:detailed_balance_for_vlt_chmc}, the detailed balance condition of VLT-CHMC was derived using the notations of Algorithm~\ref{alg:vlt_chmc}. However, it is easy to see that the same analysis carries through when we replace the reversible map $\F_\ds$ of Section~\ref{sec:vlt_chmc_special} with any reversible map as long as the set $S$ and $S^*$ satisfies \eqref{eq:generalized_reversibility}. In this section, we establish the last piece in our proof of the general VLT-CHMC algorithm; the property \eqref{eq:generalized_reversibility} holds whenever $N$ satisfies the short-return \eqref{eq:short_return} and order-preserving condition \eqref{eq:order_perserving}.
	
	We consider a generic reversible map $\F$ with an associated involution $\R$ on a general phase space $\z$ as well as a generic trajectory length function $N$ satisfying the short-return and order-preserving condition. However, all the notations and definitions directly parallel those in our presentation of the special case of VLT-CHMC in Section~\ref{sec:vlt_chmc_algorithm_in_special_case}. Fix $\z_0$ and denote $\z_i = \F^i(\z_0)$, $\z_0^* = \R \circ \F^N(\z_0)$, and $\z_i^* = \F^i(\z_0^*)$. A trajectory function $N$ determines the sets via the formula $S = \left\{\z_{- \ell}, \thinspace \ldots, \z_r \right\}$ and $S^* = \left\{\z^*_{- \ell^*}, \thinspace \ldots, \z^*_{r^*} \right\}$ where $\ell, r, \ell^*, r^* \geq 0$ are defined as
	\begin{equation} 
	\label{eq:nstep_forward_backward_general}
	\begin{aligned}
	\ell
		&= \max \left\{ i \geq 0: \F^N(\z_{- i}) = \F^N(\z_0) \right\} \\
	r
		&= \max \left\{ i \geq 0: \F^N(\z_{i}) = \F^N(\z_0) \right\} \\
	\ell^*
		&= \max \left\{ i \geq 0: \F^N(\z^*_{- i}) = \F^N(\z^*_0) \right\} \\
	r^*
		&= \max \left\{ i \geq 0: \F^N(\z^*_{i}) = \F^N(\z^*_0) \right\}
	\end{aligned}
	\end{equation}
	
	To build the intuition behind the proof, we define a partial ordering $\preceq$ on the phase space as follows:
	\begin{equation}
	\z \preceq \tilde{\z}
		\quad \text{ if } \ \F^i(\z) = \tilde{\z}
		\text{ for } i \geq 0
	\end{equation}
	Note that $\z \preceq \tilde{\z}$ if and only if $\R(\tilde{\z}) \preceq \R(\z)$, due to the reversibility of $\F$. With this notation, the short-return condition can be expressed as
	\begin{equation}
	\label{eq:short_return_ordering}
	\z \preceq \R \circ \F^N(\z^*)
		\quad \text{ for } \ \z^* = \R \circ \F^N(\z)
	\end{equation}
	The condition \eqref{eq:short_return_ordering} can be interpreted intuitively as follows; according to the trajectory termination criteria imposed by $N$, the reverse trajectory $\z^*_0, \z^*_1, \ldots$ must terminate at $\z_0$ or at $\z_i$ for $i > 0$ before coming all the way back to $\z_0$. The order-preserving condition simply amounts to
	\begin{equation}
	\F^N(\z) \preceq \F^N(\tilde{\z})
		\quad \text{ if } \ \z \preceq \tilde{\z}
	\end{equation}
	
	We now show how the order-preserving and short-return condition implies \eqref{eq:generalized_reversibility}. By the order-preserving condition, we know that
	\begin{equation}
	\F^N (\z_{- \ell}) 
		\preceq \F^N (\z_i) 
		\preceq \F^N (\z_{r})
	\end{equation}
	for all $-\ell \leq i \leq r$. On the other hand, we have $\F^N (\z_{- \ell}) = \F^N (\z_{r}) = \R(\z_0^*)$ by the definition of $\ell$ and $r$, so it follows that $\R \circ \F^N(\left\{\z_{- \ell}, \thinspace \ldots, \z_r \right\}) = \{\z_0^*\}$.  
	
	We now turn to demonstration of $\R \circ \F^N(S^*) = \{\z_r\}$. To this end, it suffices to show $\R \circ \F^N(\z_0^*) = \z_r$ as the definition of $\ell^*$ and $r^*$ combined with the order-preserving condition implies $\R \circ \F^N(\z_i^*) = \R \circ \F^N(\z_0^*)$ for all $-\ell^* \leq i \leq r^*$. Since $\R \circ \F^N(\z_r) = \z_0^*$, the short-return condition tells us $\R \circ \F^N(\z_0^*) = \z_{r+k}$ for some $k \geq 0$. To show that $k = 0$, first observe that an application of the short-return condition to the state $\z_{r+k}$ implies $\z_0^* \preceq \R \circ \F^N(\z_{r+k})$. On the other hand, the order-preserving condition implies $\F^N(\z_r) \preceq \F^N(\z_{r+k})$ and hence $\R \circ \F^N(\z_{r+k}) \preceq \R \circ \F^N(\z_r) = \z_0^*$. The preceding inequalities together show that $\R \circ \F^N(\z_{r+k}) = \z_0^*$. Since $r$ was defined as the largest integer $i$ such that $\R \circ \F^N(\z_{i}) = \z_0^*$, it follows that $k = 0$ and $\R \circ \F^N(\z_0^*) = \z_r$.

	The remaining relations in \eqref{eq:generalized_reversibility} as well as the fact $r^* = 0$ can be proved similarly with repeated applications of the short-return and order-preserving properties.

\bibliography{VLT_CHMC}

\end{document}